# Alloy formation at the bottom and top Fe interfaces of Ti/Fe/Ti trilayers

P. Süle[1], L. Bottyán[2], L. Bujdosó[2], Z. E. Horváth[1] D. Kaptás[2] A. Kovács[3], D.G. Merkel[2], A. Nakanishi[4], Sz. Sajti[2], and J. Balogh[2]

[1] Centre for Energy Research, Hungarian Academy of Sciences H-1525 Budapest 114 P.O. Box 49, Hungary
[2] Wigner Research Centre for Physics, Hungarian Academy of Sciences H-1525 Budapest 114 P.O. Box 49, Hungary
[3] Ernst Ruska-Centre for Microscopy and Spectroscopy with Electrons, Forschungszentrum Jülich, 52425 Germany
[4] Department of Physics, Shiga University of Medical Science, Shiga 520-2192, Japan

**ABSTRACT**

The chemical mixing at the Fe-on-Ti and Ti-on-Fe, i.e. the bottom and top interfaces of Fe in atomically deposited layers, were studied experimentally and by molecular dynamics simulations of the layer growth. The basic structure and chemical composition of the layers were explored in cross-sections using transmission electron microscopy. The calculations show a concentration distribution along the layer growth direction which is atomically sharp at the Ti-on-Fe interface for the (001) and (110) crystallographic orientations of the Fe layer, while it varies over a few atomic layers for Fe(111) substrate and at the Fe-on-Ti interface for all studied substrate orientations. Conversion electron Mössbauer spectroscopy and X-ray reflectometry measurements indicate an even larger asymmetry of the bottom and top Fe interface in Ti/Fe/Ti trilayers grown over Si(111) substrate by vacuum evaporation.

## I. INTRODUCTION

The special magnetic properties of surfaces and interfaces induce a variety of new magnetic phenomena [1] in nanoscale magnetic-nonmagnetic bilayers or multilayer heterostructures. The knowledge of the atomic structure and the element distribution across the interface play a key role in understanding and in technological application of such systems. Fe-Ti multilayers have been extensively studied in the past and the magnetic properties were shown to depend on the thickness of both the magnetic Fe and the nonmagnetic Ti layers [2, 3, 4] in the few nm thickness range, mainly as a consequence of interface formation.

Sequential deposition of Fe and Ti by sputtering [2, 3, 4] results in chemical mixing at the interface of body centered cubic (bcc) Fe and hexagonal close packed (hcp) Ti layers. The interface may contain bcc-$Fe_{100-x}Ti_x$ (for x<10), $Fe_2Ti$, FeTi, hcp-$Ti_{100-y}Fe_y$ (for y<1) crystalline and also various amorphous alloys similar to those obtained by co-sputtering [5, 6]. The phase formation under ion-irradiation [7] or heat treatment [8, 9] of multilayers has also been investigated, but the Ti-on-Fe and the Fe-on-Ti interfaces were assumed equivalent in all these multilayer studies. Trilayers or single interfaces have rarely been investigated [10, 11]. An elaborate high-resolution transmission electron microscopy (TEM) study of multilayer samples was interpreted in terms of interface asymmetry, i.e. non-equivalent Fe-on-Ti and Ti-on-Fe interfaces [12], but the interpretation was based on model calculations. Upcoming TEM investigations [3, 9, 13, 14] gave no evidence for any interface asymmetry and the results on the phase formation were also controversial. Beyond the experimental studies molecular dynamics (MD) simulations were performed for the case when Ti was deposited onto Fe(001) substrate [15]. The results were indicative of an almost atomically sharp Ti-on-Fe interface under room temperature (RT) growth. These controversial results necessitate a more elaborate experimental and theoretical work in order to reveal the interface properties of polycrystalline Fe-Ti multilayers.

The atomistic simulation of the interface formation during the atomic deposition process is a feasible theoretical approach up to about 100 $nm^2$ surfaces. The experimental methods most frequently used to characterize the multilayer interfaces, e.g. grazing incidence x-ray or neutron reflectivity, Rutherford backscattering or Auger electron spectroscopy, etc., investigate samples of macroscopic size and they are unable to make clear distinction between the topological roughness / waviness of the interface and the chemical mixing of the layers. The various TEM methods may give atomic scale information about both the topological features and the chemical mixing, but they sample only a nanoscale section of a macroscopic sample and the sample processing might change the structure in some cases. Consequently, a quantitative comparison of the experimental results and the MD simulation is not unambiguous. Mössbauer spectroscopy (MS), being sensitive to the atomic scale neighborhood of a specific isotope, can give unique information on the extent of the chemical mixing and on the nature of the compound phases formed at the interface. However, additional information on the morphology of the sample is inevitable for a sound identification of the different Mössbauer spectral components. For example, [57]Fe probe nuclei surrounded exclusively by Fe atoms in the first and second coordination shell in a body centered cubic structure experience a hyperfine field similar to that in pure bcc-Fe, no matter if they sit at the interface of an iron layer, inside a small iron cluster or in a random dilute alloy.

In this work we study the alloy formation at the Ti-on-Fe and Fe-on-Ti interfaces in Ti/Fe/Ti trilayers by applying Mössbauer spectroscopy complemented by cross-sectional TEM and X-ray reflectometry (XRR). To attain the best signal to noise ratio conversion electron Mössbauer spectroscopy (CEMS) was applied. Samples were prepared using iron metal highly enriched in the $^{57}$Fe Mössbauer resonant isotope. In order to determine the phase fractions within the top (Ti-on-Fe) and bottom (Fe-on-Ti) Fe interface Ti/$^{57}$Fe/Ti and Ag/$^{57}$Fe/Ti sample pairs were compared exploiting the non-mixing property [16, 17] of Fe and Ag. Analyzing the spectra of Ag/$^{57}$Fe/Ti samples provides information on the Fe-on-Ti interface, since the Ag/Fe interface is chemically sharp and the Ag layer causes only a small and well documented change [18] in the hyperfine parameters of $^{57}$Fe within the atomic layers nearest and next-nearest to the Fe-Ag interface. Additional spectral components appearing in the Mössbauer spectra of Ti/$^{57}$Fe/Ti samples reveal the specific properties of the Ti-on-Fe interface. The experimental results are compared with MD simulations of the layer growth performed for all basic crystal orientations of the respective substrate layer, Ti or Fe. The samples were prepared by evaporation in high vacuum, which, to our knowledge, has not yet been used in the interface studies of this system.

## II. EXPERIMENTAL METHODS

The metallic layers were prepared by vacuum evaporation onto Si(111) substrates. The parameters of the evaporation chamber are described in Ref. 19. Ti and Ag were evaporated by electron beam, while the iron (95 per cent enriched in the $^{57}$Fe isotope) was evaporated by resistive heating from a tungsten crucible. Three sample pairs with the following layer sequences were studied:

| | |
|---|---|
| 20nm Ti / 1.5nm $^{57}$Fe / 10nm Ti / Si | (1A) |
| 10nm Ti / 10nm Ag / 1.5nm $^{57}$Fe / 10nm Ti / Si | (1B) |
| 20nm Ti / 2.5nm $^{57}$Fe / 10nm Ti / Si | (2A) |
| 10nm Ti / 10nm Ag / 2.5nm $^{57}$Fe / 10nm Ti / Si | (2B) |
| 20nm Ti / 3.5nm $^{57}$Fe / 20nm Ti / Si | (3A) |
| 10nm Ti / 10nm Ag / 3.5nm $^{57}$Fe / 20nm Ti / Si | (3B) |

The $^{57}$Fe layers of samples A and B were deposited simultaneously to ensure the equal thickness of the $^{57}$Fe layers of the sample pairs. The tungsten crucible used for $^{57}$Fe was placed at equal distance from the two substrates. Similarly, the respective Ti layers of the sample pairs were deposited simultaneously, but the distance to the substrates was somewhat different, which may result in some (below 1 nm) difference in the Ti layer thickness of the members of the pairs. The Ti layers covering the Ag layers in samples 1B, 2B and 3B serve as capping to minimize oxidation of the $^{57}$Fe, since thin Ag layers may get easily fractured. Hereafter when necessary for easier reading the A and B samples will be denoted as Ti/x Fe/Ti and Ag/x Fe/Ti, respectively, where x will give (in nm units) the thickness of the $^{57}$Fe layer of the sample in question.

Structure characterization has been performed on sample Ti/3.5 Fe/Ti. Specimens for cross-sectional studies were prepared in a dual-beam scanning electron microscopy and Ga$^+$ focused ion beam (FIB) system (FEI Helios 440), which allowed to keep the sample temperature unaltered during the preparation. Pure carbon was deposited on the surface as protective layer. The surface damage induced by the high-energy Ga beam was reduced by low-energy Ar$^+$ ion beam milling using Fischione NanoMill 1040 system at energies below 1 keV. The structure and chemical composition of the trilayer was studied using a conventional TEM (FEI Tecnai G2) and a probe-aberration corrected scanning TEM (FEI Titan G2 80-200) equipped with in-column energy-dispersive X-ray spectroscopy (EDXS) detectors [20]. High-angle annular dark-field (HAADF) scanning TEM images were recorded at 200 kV using the detector convergence semi-angle of 69.1 mrad. The high-resolution images and elemental maps were processed using the FEI Velox software.

As an alternative to TEM the layer structure of sample Ti/3.5 Fe/Ti, was studied by specular X-ray reflectivity measurements, too. XRR probes the refractive index depth profile in flat samples perpendicular to the surface. Interpretation of the reflectivity involves modeling this depth profile by an approximate multilayer structure of individual constant refractive indices (determined by the composition and density) and interface regions in between. XR measurements were performed by a Bruker AXS D8 Discover diffractometer equipped with Göbel-mirror and a scintillation detector using Cu Kα radiation. The collected reflectivity data were evaluated by using the FitSuite code [21], a free multipurpose software. FitSuite uses a transfer matrix method described in Ref. 22 for calculation of X-ray reflectograms. It treats the interface roughness as described in Refs. 23 and 24 assuming and error function-type interface depth profile characterized by its standard deviation, the (rms) roughness. This method is unable to distinguish the interface roughness and laterally homogeneous intermixing e.g. due to interdiffusion.

The conversion electron Mössbauer spectroscopy measurements were carried out by a conventional constant acceleration spectrometer. For the detection of the conversion electrons low background gas filled proportional counter was used with pure $H_2$ gas at low temperatures [25], and 96%He-4%$CH_4$ gas mixture at room temperature. The spectra were measured by a 50 mCi $^{57}$Co(Rh) single line Mössbauer source. The hyperfine field (HF) distributions were evaluated according to the Hesse-Rübartsch method [26], by fitting the amplitudes of a number of sextets with HFs increasing with equal step values. The isomer shift (IS) values are given relative to that of α-Fe at room temperature.

## III. ATOMISTIC SIMULATION METHODS

Classical molecular dynamics was used as implemented in the LAMMPS code (Large-scale Atomic/Molecular Massively Parallel Simulator) [27] in parallel environment using GPU nodes reaching a significant speedup in the calculations. Isobaric-isothermal (NPT ensemble) simulations (with Nose-Hoover thermostat and a prestostat) were carried out at 300 K, allowing for the variation of the supercell size and shape which ensured further flexibility in absorbing pressure waves and/or in relaxing the occasional strain in the system. Periodic boundary conditions were ensured along the lateral directions (x,y) and vacuum regions were inserted above and below the slab of the film/substrate system. The vacuum region was chosen sufficiently thick (a few nm) to keep sufficient space between the evaporated particles and the forming surface of the deposit. The temperature of the substrate was kept at 300 K and the variable time step algorithm was exploited. The atomic and nanoscale structures were displayed using the OVITO code [28].

The specific (fix deposit) algorithm as implemented in LAMMPS was used for the simulation of thin film growth. The flux of the deposited atoms was kept at 2.25*$10^{25}$ atoms/s·$cm^2$ (one particle per 10k simulation steps). In total, at least 15-20 monolayers (ML) of adatoms were deposited on the substrate in which each ML corresponds to nearly 200 adatoms for the Fe and Ti layers on the 4 x 4 $nm^2$ substrate surface. The EAM potentials were used for both Fe and Ti as generated by the EAM database tool [29] provided by the code LAMMPS. For the FeTi cross-interaction the mixing rule of Johnson [30] was applied which provides reasonable alloy properties. Further details of the simulated thin film growth can be found in Ref. 17.

## IV. EXPERIMENTAL RESULTS

### A. Electron microscopy

The structure of the deposited layers of sample Ti/3.5 Fe/Ti was studied in cross-section using various TEM methods, as shown in Figure 1. Figure 1a displays a bright-field STEM image of the Ti and Fe layers deposited on Si substrate. Both Ti layers show polycrystalline structure with grain sizes ranging from 5 to 15 nm. The light contrast of the top of the upper Ti layer is due to surface oxidation. The roughness of the Ti and Fe surfaces is estimated to be around 1 nm. Figure 1b shows an atomic resolution HAADF STEM image of the Ti/Fe/Ti layers. For the given imaging conditions, the HAADF STEM image contrast scales approximately with the atomic number as I ~ $Z^{1.7}$ which results in the bright (dark) contrast of the Fe (Ti) layer. The Fe layer thickness on the STEM images was measured to be 4.2 nm that is slightly larger than the nominal thickness. The digital diffractogram of a high-resolution image in Fig. 1c shows the Ti and Fe lattice arrangements relative to each other. The analysis of the pattern suggests [010] orientation of Ti and [100] orientation of Fe. The lattice distances across the Ti-on-Fe and the Fe-on-Ti interfaces, were measured on the HAADF STEM image by taking integrated intensity scans across the regions marked by boxes in Fig. 1b and plotted in Fig. 1d. The dashed lines at 0.234 nm and at 0.207 nm correspond to the Ti (002) and Fe (110) lattice planes, respectively. The transition at the Ti-on-Fe interface seems to be abrupt, while at the Fe-on-Ti interface one intermediate plane is present. Fig. 1e shows the color coded chemical composition map of the Ti and Fe layers extracted from EDXS recordings. The standard background removal and the Cliff-Lorimer method were applied to calculate the atomic fraction of the elements. Fig. 1f shows the corresponding variation across the interfaces measured in the map of Fig. 1e. The Fe concentration is approximately 0.94 in the middle of the Fe layer. EDXS does not reveal differences between the Fe-on-Ti and the Ti-on-Fe transition zones.

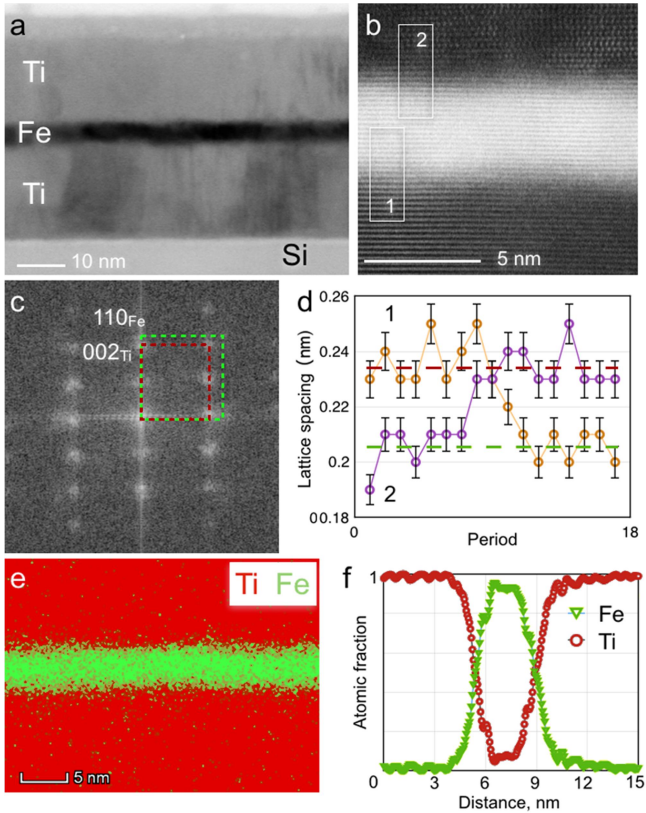

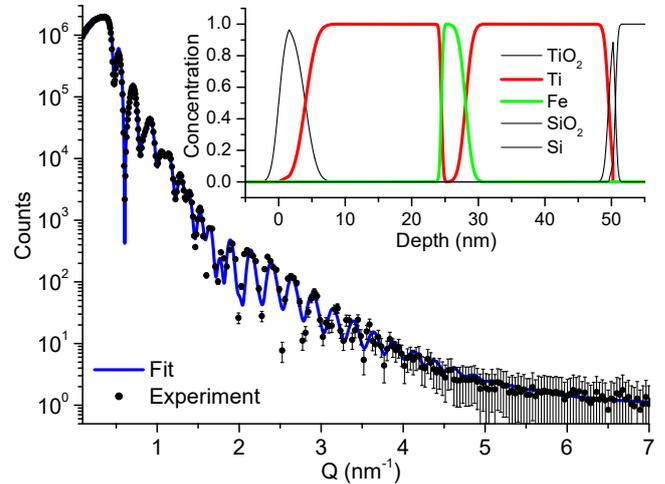

FIG. 1. Structure of the Ti/3.5 Fe/Ti sample studied in cross-section using various TEM methods; bright-field STEM image of the Ti and Fe layers deposited on Si substrate (a), atomic resolution HAADF STEM image (b), digital diffractogram of the high-resolution image (c), lattice distances across the regions marked by boxes on the HAADF STEM image (d), color coded chemical composition map of the Ti and Fe layers extracted from EDXS recordings (e), atomic fraction of the elements across the interfaces calculated from the composition map (f).

FIG. 2. Measured and fitted X-ray reflectivity curve for the Ti/3.5nm Fe/Ti sample. The inset shows the concentration profiles of the different materials corresponding to the fitted parameters. Red (dark) and green (light) thick lines indicate Ti and Fe. The components appearing around 0 and 50 nm depth and above - indicated by thin black lines - belong to $TiO_x$, $SiO_2$, and Si, respectively. The asymmetry of the Fe-Ti interface is salient.

## B. X-ray reflectivity

The measurements were performed on the Ti/3.5 Fe/Ti sample using the Cu Kα line (0.154 nm). The measured and fitted curves and the fitted parameters are shown in Fig. 2 and Table I, respectively. The layer thicknesses were found to be 21.5 nm, 3.63 nm, 20.4 and 4 nm for the bottom Ti, Fe, top Ti, and $TiO_x$ layers, respectively. Significant intermixing occurs only at the bottom of the Fe layer, i.e. at the Fe-on-Ti interface, as the roughness of the bottom Ti layer is 0.86 nm while that of the Fe layer is found to be quite sharp 0.23 nm. A native 1 nm $SiO_2$ layer on the surface of the Si substrate is also included in the layer model, according the TEM results.

TABLE I. The parameters obtained from the fit of x-ray reflectometry measurements, the roughness values refer to the top interface of the corresponding layer. The refractive indices not shown here were obtained from the webpage http://henke.lbl.gov/optical_constants/getdb2.html provided by Eric Gullikson based on Ref. 31. The densities of the $TiO_x$ and $SiO_2$ were assumed to be 3.78 and 2.2 g·cm$^{-3}$, respectively. These densities correspond to anatase and amorphous $SiO_2$.

|                | $TiO_x$      | top Ti        | $^{57}Fe$     | bottom Ti     | $SiO_2$     |
|----------------|--------------|---------------|---------------|---------------|-------------|
| Thickness (nm) | 4.0 ± 0.02   | 20.4 ± 0.02   | 3.63 ± 0.02   | 21.5 ± 0.04   | 1 ± 0.03    |
| Roughness (nm) | 0.87 ± 0.02  | 1.36 ± 0.01   | 0.23 ± 0.05   | 0.86 ± 0.03   | 0.53 ± 0.01 |

## C. Mössbauer spectroscopy

The CEMS spectra of the three as-received sample pairs measured at RT and 15K are shown in Fig. 3a and b, respectively. All spectra (except that of sample 1A at RT) were fitted by a set of sextets with hyperfine fields allowed in the 5-40T range and a broadened doublet with a single quadrupole splitting (QS). The evaluated normalized HF distributions are shown as insets to each spectrum in Fig. 3. In the spectrum fitting the line-width of the sextet components of the HF distribution was fixed to 0.24mm/s, the step value was an iteration parameter in the 0.7-0.8T range. The isomer shifts were assumed proportional to the HFs, the lower the HF the larger negative value to have. The intensities of the second and fifth lines relative to that of line three and four of the sextets were close to 4 in each case, indicating an in-plane orientation of the magnetic moments.

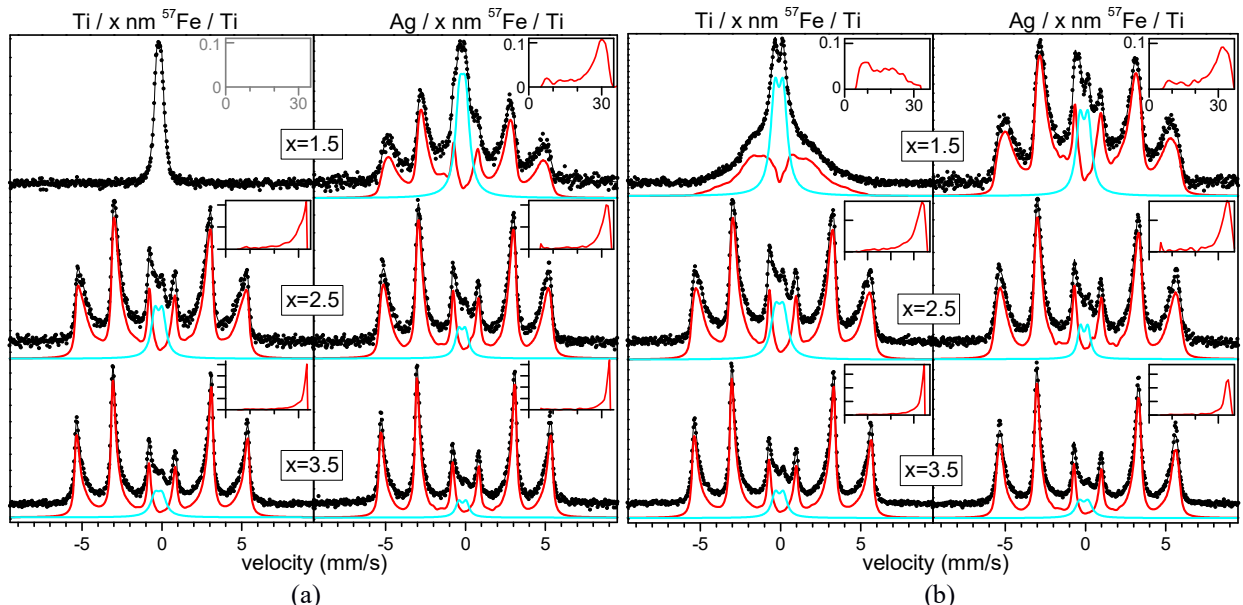

FIG. 3. Mössbauer spectra of the as received samples measured at RT (a) and at 15 K (b). The sub-spectra belonging to the doublet component and to the HF distributions (shown in the insets) are indicated by blue (light) and red (dark) lines, respectively. In the insets the horizontal (HF) scales are the same for all HF distributions and are labelled in the topmost insets. The vertical intensity scales of the insets, however, are normalized to unit area and vary from graph to graph one tick belonging to p(B)*B=0.1 in each inset.

TABLE II. Mössbauer parameters of the spectra measured at RT and 15K; Isomer shift (IS), quadrupole splitting (QS) and linewidth (W) of the doublet, average hyperfine field of the magnetic alloy component ($HF_{alloy}$) and the iron content of the different components in layer thickness equivalent (see text for explanation) as calculated from the spectral intensities and the nominal thickness of deposited Fe layer. Typical errors are given in the second row.

|  | IS (mm/s) | QS (mm/s) | W (mm/s) | $HF_{alloy}$ (T) | $d_{bcc}$ (nm) | $d_{para}$ (nm) | $d_{mag}$ (nm) |
|---|---|---|---|---|---|---|---|
|  | ±0.02 | ±0.02 | ±0.05 | ±0.5 | ±0.05 | ±0.05 | ±0.05 |
|  | RT/15K | RT/15K | RT/15K | RT/15K | RT/15K | RT/15K | RT/15K |
| Ti/1.5 Fe/Ti | -0.17/-0.10 | 0.29/0.52 | 0.40/0.62 | 0/16.4 | 0/0 | 1.50/0.66 | 0.00/0.84 |
| Ag/1.5 Fe/Ti | -0.18/-0.10 | 0.38/0.51 | 0.61/0.57 | 25.0/25.2 | 0.04/0.25 | 0.46/0.24 | 1.00/1.01 |
| Ti/2.5 Fe/Ti | -0.21/-0.07 | 0.47/0.50 | 0.56/0.69 | 26.3/27.5 | 0.61/0.66 | 0.33/0.39 | 1.56/1.45 |
| Ag/2.5 Fe/Ti | -0.22/-0.09 | 0.42/0.46 | 0.51/0.43 | 27.1/27.0 | 0.54/0.91 | 0.21/0.19 | 1.75/1.40 |
| Ti/3.5 Fe/Ti | -0.21/-0.08 | 0.45/0.50 | 0.65/0.56 | 27.1/27.5 | 1.53/1.46 | 0.35/0.33 | 1.62/1.71 |
| Ag/3.5 Fe/Ti | -0.19/-0.11 | 0.46/0.54 | 0.47/0.61 | 27.0/28.3 | 1.73/1.80 | 0.21/0.26 | 1.56/1.44 |

The most important parameters of the spectra are summarized in Table II. The magnetic component is divided into two subgroups for each spectrum. The one indicated as bcc-Fe, is a sum of the sextets in the 0.5T vicinity of 33 and 33.8T, the literature HF values of bcc-Fe at RT and 15K, respectively. This component can be attributed to Fe atoms without Ti or Ag as first neighbors. The component stemming from the sextet intensities below 32.5 and 33.3T at RT and 15K, respectively is labeled as magnetic alloy and belongs to Fe atoms with Ti (or Ag) neighbors in the first and/or second neighbor shells [32]. The average HF of this component is largely different for samples 1A and 1B, but remains constant within the ~0.5T experimental error for x=2.5 and 3.5 nm. The spectral fraction of the different components was transformed to Fe layer thickness equivalent by calculating the product of the spectral fractions and the nominal thickness of the Fe layer. (The same

Mössbauer-Lamb factor was assumed for all components.) These values are labeled as $d_{bcc}$, $d_{para}$, and $d_{mag}$ for the bcc-Fe, paramagnetic alloy and magnetic alloy components, respectively.

## V. RESULTS OF ATOMISTIC SIMULATIONS

Classical MD simulations have been carried out for the thin film growth of Fe-Ti bilayers with different orientations of either the Fe or the Ti substrate. The cross-sectional view of the grown interfaces is shown in Figs 4a and 4b and the calculated concentration depth profiles for different substrate orientations are displayed in Fig. 5. The Fe-on-Ti interfaces are slightly intermixed and the interface broadening depends on the orientation of the substrate, although the variation is small. The width of the interface width is around 1 nm for Fe/Ti(1101) and 0.8 nm for Fe/Ti(0001). In the Ti-on-Fe case no atomic mixing can be seen for the Fe(001) substrate orientation, but the interface is slightly wavy, as can be seen in Fig. 4b, which results in a 0.3nm intermixing in the concentration depth profile. For the Fe(110) orientation a modest intermixing also takes place and the intermixed region is slightly larger, 0.5nm. The largest intermixing takes place in case of the Fe(111) substrate orientation, but the interface width (0.7nm) still remains below the Fe-on-Ti values. Altogether, from the simulations we can conclude that the intermixing is asymmetric with respect to the interchange of the constituents of the film and the substrate.

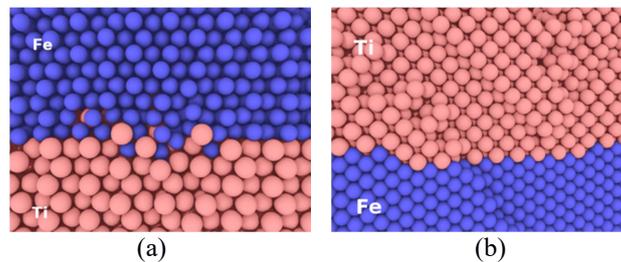

(a)          (b)

FIG. 4. Typical cross-sectional view with 1 nm slab thickness of Fe/Ti(0001) (a) and Ti/Fe(001) (b) interfaces as obtained from MD simulations. Blue (dark) and red (light) spheres are Fe and Ti atoms, respectively.

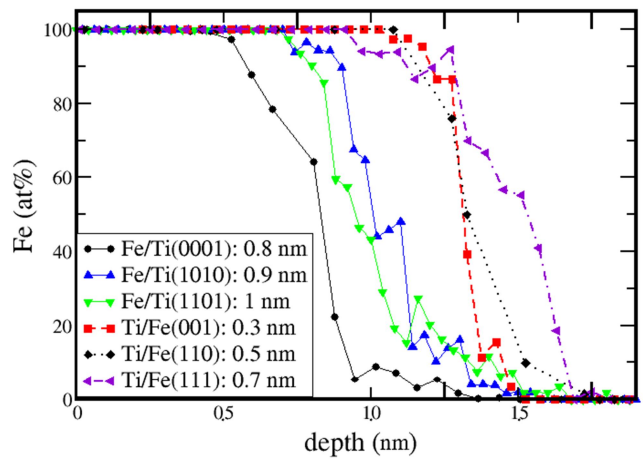

FIG. 5. Concentration profiles across the Fe-on-Ti and Ti-on-Fe interfaces as calculated from MD simulations for different orientations of the substrate layer. The substrate orientations which belong to the different curves and the thickness of the respective interface alloy are indicated in the inset.

## VI. DISCUSSION

The most direct information on the range of the chemical mixing is obtained from Mössbauer spectroscopy, therefore first we discuss those results. The room temperature Mössbauer spectra of the Ti/x Fe/Ti samples shown in Fig. 3a are in line with previous multilayer studies [2, 4]. Below the 2nm thickness of the deposited Fe layer all Fe atoms are alloyed with Ti in an amorphous structure as suggested by the observed room temperature IS and QS values, -0.18mm/s and 0.38mm/s, respectively. The low temperature measurement, however, reveals that the amorphous phase is not homogeneous; about 40% of the Fe atoms remain paramagnetic, while the rest exhibits a broad HF distribution in the 5T-30T range. The Fe-on-Ti and Ti-on-Fe interfaces certainly overlap in the Ti/1.5 Fe/Ti sample and consequently the magnetic component, which is well resolved in the RT spectrum of Ag/1.5 Fe/Ti, is not present in the RT spectrum of Ti/1.5 Fe/Ti. The single Fe-on-Ti interface of the Ag/1.5 Fe/Ti sample, beside amorphous alloys, contains crystalline bcc-$Fe_{100-x}Ti_x$ alloy phase, as well, since the critical temperature of all other crystalline phases and of the amorphous alloys as well is below RT for the entire concentration range.

Interestingly, the Ag/1.5 Fe/Ti spectra exhibit HF components around 10T at room temperature and at 15K alike, which was neither observed in high Ti content crystalline [33] nor in sputtered amorphous Fe-Ti alloys [34]. The influence of the Ag layer can be suspected first, but the equilibrium solubility of Ag is very low in both Fe, and in Fe-Ti crystalline alloys [35]. Moreover, the Ag layer causes only a minute decrease/increase of the hyperfine field [18] in pure bcc-Fe layers at RT/15K, an effect restricted to 2-3 monolayers (MLs) near the interface. Additionally, the low-field components are also present in samples Ti/2.5 Fe/Ti and Ag/2.5 Fe/Ti. (In the case of Ti/3.5 Fe/Ti and Ag/3.5 Fe/Ti the corresponding intensities are far too small to exceed the experimental error.) With these arguments the 10T spectral components is attributed to such non-magnetic Fe atoms which have magnetic Fe neighbours and exhibit magnetic splitting due to the hyperfine field transferred via the spin polarized conduction electrons [36]. Such Fe atoms are probably present in the thin paramagnetic amorphous layer, but, due to their small fraction (corresponding to 0.2 nm Fe), mostly remain hidden [2, 4] under the magnetic alloy contribution.

The interface components ($d_{para} + d_{mag}$) make about 2nm Fe layer equivalent in samples Ti/2.5 Fe/Ti and Ti/3.5 Fe/Ti, which is supported by former literature results [2, 4] and by the observation that the Fe-on-Ti and Ti-on-Fe interfaces overlap in Ti/1.5 Fe/Ti. The bcc-Fe component ($d_{bcc}$) is 0.64nm and 1.50nm, respectively for the Ti/2.5 Fe/Ti and the Ti/3.5 Fe/Ti samples. Although the bcc-Fe component might as well belong to Fe atoms without Ti neighbours in a dilute alloy, the formation of a continuous bcc-Fe layer in the Ti/3.5 Fe/Ti sample seems to be justified by several observations, such as *i)* $d_{bcc}$ increases 0.86nm for 1nm increase of the nominally deposited Fe thickness, *ii)* the interface properties ($d_{para}$, $d_{mag}$, QS, $HF_{alloy}$) do not depend on the deposited Fe thickness indicating that the Fe-on-Ti and Ti-on-Fe interfaces do not overlap. As it will be shown, the XRR also supports the presence of a continuous Fe layer in the Ti/3.5 Fe/Ti sample. In view of a continuous Fe layer separating the Fe-on-Ti and Ti-on-Fe interfaces, the minor difference between the spectra of Ti/3.5 Fe/Ti and Ag/3.5 Fe/Ti samples (*c.f.* Fig. 3) means that the Ti-on-Fe interface is rather similar to the Ag-on-Fe interface, namely chemically rather sharp. This result is well supported by the MD calculations.

In the following, from $d_{para}$ and $d_{mag}$ the relative phase volumes / partial thicknesses will be estimated by considering the known concentration limits of the identified amorphous and crystalline alloys. In lack of experimental data on the density of the different compounds, the calculations are based on bulk densities of the elements in appropriate ratios according to the atomic concentrations. Finally the derived layer and interface thicknesses will be compared to the TEM and XRR results. Since the amorphous alloy and the crystalline bcc alloy phases are clearly distinguished at RT, the interface widths will be calculated from the RT spectra.

Comparing the Ti/3.5 Fe/Ti and Ag/3.5 Fe/Ti spectra helps to unravel details about the rather sharp Ti-on-Fe interface. In Ti/3.5 Fe/Ti $d_{para}$ and $d_{mag}$ are slightly higher on the expense of $d_{bcc}$, while the hyperfine parameters (IS, QS and $HF_{alloy}$) are practically the same in the two samples. This indicates the presence of similar phases in different amounts near the corresponding interfaces. The small difference in the intensity of the paramagnetic component is well seen in the spectra of Fig. 3. Accordingly, $d_{para}$ is 0.14nm larger for Ti/3.5 Fe/Ti than for Ag/3.5 Fe/Ti. The slight difference in the magnetic alloy phase contribution is not well seen in the spectra, but from the evaluations $d_{mag}$ is also larger for Ti/3.5 Fe/Ti than for Ag/3.5 Fe/Ti, by 0.06nm. Here one has to take into account that Ag/3.5 Fe/Ti contains two monolayer Fe (~ 0.3nm), whose hyperfine field is modified by the Ag layer. This means that $d_{mag}$ has to be decreased by 0.3nm for Ag/3.5 Fe/Ti and by that the difference becomes 0.36 nm. Assuming the interface to form a continuous layer, its thickness can be estimated from the Fe content and the alloy composition of the two interface components. With the $Fe_{30}Ti_{70}$ and $Fe_{80}Ti_{20}$ hypothetical concentrations of the amorphous and the crystalline alloys, the corresponding widths are 0.63nm and 0.49nm, respectively. The sum of the two, 1.12nm gives an upper limit of the Ti-on-Fe interface width, since the maximum possible Ti concentrations of the amorphous [37] and the bcc alloys [33] were used in the calculation and lower Ti concentrations would result in smaller thickness. The temperature dependences clearly indicate a concentration distribution; therefore the average Ti concentrations are certainly lower than the maximum possible 70 and 20 at%. If 50 and 10 at% average concentrations are assumed for the amorphous and the crystalline alloys the Ti-on-Fe interface width comes to 0.77 nm. Additionally, the Ti-on-Fe interface is most probably not a completely continuous layer.

The Fe-on-Ti interface, as deduced from the parameters of the Ag/3.5 Fe/Ti sample, is more extended than the Ti-on-Fe interface. The paramagnetic component contains 0.21nm Fe, which forms a layer of 0.52nm thickness assuming a composition of $Fe_{50}Ti_{50}$. The magnetic alloy component is much thicker, $d_{mag}$=1.56nm, which gives a 1.47nm thick bcc alloy layer after the 0.3 nm correction for the Ag/Fe interface and supposing a $Fe_{90}Ti_{10}$ composition. After summing up the thicknesses calculated from $d_{para}$ and $d_{mag}$, one can say that if 50 and 10 at% average concentrations are supposed for the amorphous and the crystalline alloys, the Fe-on-Ti interface thickness extends to 1.99 nm.

From the above Mössbauer results one can conclude that the bottom and the top interface of a Fe layer in between Ti layers are very different both in the extent and in the ratio of the amorphous and the crystalline alloys appearing at the interface. The

Fe-on-Ti bottom interface is almost three times thicker than the Ti-on-Fe top interface and the ratio of the Fe-rich bbc alloy is larger than the Ti-rich amorphous alloy. The Ti-rich amorphous alloy has a higher ratio than the crystalline alloy in the thin Ti-on-Fe interface, which probably does not form a continuous layer.

The XRR measurements also indicate a significant asymmetry of the Fe-on-Ti and Ti-on-Fe interfaces. Defining the interface width as the full width at half maximum $2\sqrt{2\ln 2}\,\sigma \approx 2.355\,\sigma$ of the corresponding normal distribution, where σ is the interface (rms) roughness, roughness values of 0.86 nm (Ti) and 0.23 nm (Fe) result in 2.0 and 0.54 nm "10-90" interface widths, respectively. These values agree quite well with the values estimated from the Mössbauer results (1.99nm and 0.77nm). The scaling of the interface width may be model dependent, but a strong asymmetry of the Ti-on-Fe and the Fe-on-Ti interface widths is undoubtedly experimentally verified by the XRR data, too.

Some structural features, the thickness of the Ti, $TiO_x$ and $SiO_2$ that can be seen in the TEM measurements in Fig. 1a agree well with those deduced from the XRR measurements. The thickness of the middle Fe layer is, however, smaller than the 4.0 nm – 4.2 nm value which can be estimated from the dark contrast region in Fig. 1a. The deduced concentration depth profile of Fe also disagrees with that shown in Fig. 1f. Surprisingly, the TEM experiments did not show the marked asymmetry of the Fe-on-Ti and Ti-on-Fe interfaces, observed by the other two experimental methods. In our understanding, it can be due to the averaging of the signal across the specimen thickness in the view direction and a possible modest heating in spite of all the care taken during the sample thinning. To understand the missing Fe concentration and the identical interfaces in Fig. 1f, one has to consider the specimen geometry that was used for the TEM experiments. The specimen is approximately 40 nm thick and the measurement shows the projected signal in the maps. Since the grain size of Fe (and Ti) is much less than the thickness of the layer, the signal at a given point is coming from the sum of some grains and grain boundaries along the electron beam direction and the surface roughness of the layers is also averaged, resulting in a broad elemental distribution. The small crystallite size might also explain that the Fe concentration peaks at 95 at%, since the Ti deposition on Fe may results in some Ti diffusion along the grain boundaries, which appears as the presence of Ti in the middle of the Fe layer, as shown in Fig. 1f. (Diffusion along the grain boundaries might also explain that the experimentally observed interface thickness is about an order of magnitude larger than the one obtained by MD simulations supposing perfect single crystal substrate layers.) One may conclude that the chemical measurement across small-sized crystalline layers using TEM may not give clear view on the distribution of the elements. The analysis of the lattice planes measured on high-resolution TEM images (as in Fig. 1d), can be more representative, however it represents very bad statistics considering the large number of crystals measured by the other two experimental methods.

## VII. CONCLUSION

Molecular dynamics simulations of the layer growth show a concentration distribution along the layer growth direction which is atomically sharp at the Ti-on-Fe interface for the (001) and (110) crystallographic orientations of the Fe layer, while it varies over a few atomic layers for Fe(111) substrate and at the Fe-on-Ti interface for all basic crystallographic orientations of Ti. Conversion electron Mössbauer spectroscopy and X-ray reflectometry measurements are indicative of larger values of the interface widths, but support the asymmetry in the width of the bottom and top Fe interface as measured in Ti/Fe/Ti trilayers grown over Si(111) substrate by vacuum evaporation. The significantly larger experimental interface width is probably due to the polycrystalline nature of the layers and the small size of the Fe crystallites within the layers.

## ACKNOWLEDGEMENT

The authors acknowledge financial support of the Hungarian Scientific Research Fund (OTKA) Grant K112811.